# Proposal of Appropriate Location Calculations for Environment Adaptation

Yoji Yamato[a,*]

[a]Network Service Systems Laboratories, NTT Corporation, 3-9-11 Midori-cho, Musashino-shi, Tokyo 180-8585, Japan

*Corresponding Author: yoji.yamato.wa@hco.ntt.co.jp

## Abstract

To use heterogeneous hardware, programmers must have sufficient technical skills to utilize OpenMP, CUDA, and OpenCL. On the basis of this, I have proposed environment-adaptive software that enables automatic conversion, configuration, and high performance operation of once written code, in accordance with the hardware. However, although it has been considered to convert the code according to the offload devices, there has been no study where to place the offloaded applications to satisfy users' requirements of price and response time. In this paper, as a new element of environment-adapted software, I examine a method to calculate appropriate locations using linear programming method. I confirm that applications can be arranged appropriately through simulation experiments when some conditions such as application type and users' requirements are changed.

## 1. Introduction

As Moore's Law slows down, a central processing unit's (CPU's) transistor density cannot be expected to double every 1.5 years. To compensate for this, more systems are using heterogeneous hardware, such as graphics processing units (GPUs), field-programmable gate arrays (FPGAs), and many-core CPUs. For example, Microsoft's search engine Bing uses FPGAs [1], and Amazon Web Services (AWS) [2] provides GPU and FPGA instances using cloud technologies (e.g., [3]-[10]). Systems with Internet of Things (IoT) devices are also increasing (e.g., [11]-[19]).

However, to properly utilize devices other than small-core CPUs in these systems, configurations and programs must be made that consider device characteristics, such as Open Multi-Processing (OpenMP) [20], Open Computing Language (OpenCL) [21], and Compute Unified Device Architecture (CUDA) [22]. In addition, embedded software skills are needed for IoT devices detail controls. Therefore, for most programmers, skill barriers are high.

The expectations for applications using heterogeneous hardware are becoming higher; however, the skill hurdles are currently high for using them. To surmount these hurdles, application programmers should only need to write logics to be processed, and then software should adapt to the environments with heterogeneous hardware to make it easy to use such hardware. Java [23], which appeared in 1995, caused a paradigm shift in environment adaptation that allows software written once to run on another CPU machine. However, no consideration was given to the application performance at the porting destination.

Therefore, I previously proposed environment-adaptive software that effectively runs once-written applications by automatically executing code conversion and configurations so that GPUs, FPGAs, many-core CPUs, and so on can be appropriately used in deployment environments. For an elemental technology for environment-adaptive software, I also proposed a method for automatically offloading loop statements and function blocks of applications to GPUs or FPGAs [24]-[28].

This paper is for optimizing the applications placement when a normal CPU program is offloaded to a device such as a GPU, to meet the user's cost requirements and the response time requirements. I propose a method to calculate appropriate locations using linear programming method. I confirm that applications can be arranged appropriately through simulation experiments when some conditions such as application type and users' requirements are changed.

## 2. Existing Technologies

### 2.1 Technologies on the market

Java is one example of environment-adaptive software. In Java, using a virtual execution environment called Java Virtual Machine, written software can run even on machines that use different operating systems (OSes) without more compiling (Write Once, Run Anywhere). However, whether the expected performance could be attained at the porting destination was not considered, and there was too much effort involved in performance tuning and debugging at the porting destination (Write Once, Debug Everywhere).

CUDA is a major development environment for general purpose GPUs (GPGPUs) (e.g., [29]) that use GPU computational power for more than just graphics processing. To control heterogeneous hardware uniformly, the OpenCL specification and its software development kit (SDK) are widely used. CUDA and OpenCL require not only C language extension but also additional descriptions such as memory copy between GPU or FPGA devices and CPUs. Because of these programming difficulties, there are few CUDA and OpenCL programmers.

For easy heterogeneous hardware programming, there are technologies that specify parallel processing areas by specified directives, and compilers transform these specified parts into device-oriented codes on the basis of directives. Open accelerators (OpenACC) [30] and OpenMP are examples of directive-based specifications, and the Portland Group Inc. (PGI) compiler [31] and gcc are examples of compilers that support these directives.

In this way, CUDA, OpenCL, OpenACC, OpenMP, and others support GPU, FPGA, or many-core CPU offload processing. Although processing on devices can be done, sufficient application performance is difficult to attain. For example, when users use an automatic parallelization technology, such as the Intel compiler [32] for multi core CPUs, possible areas of parallel processing such as "for" loop statements are extracted. However, naive parallel execution performances with devices are not high because of overheads of CPU and device memory data transfer. To achieve high application performance with devices, CUDA, OpenCL, or so on needs to be tuned by highly skilled programmers, or an appropriate offloading area needs to be searched for by using the OpenACC compiler or other technologies. As an effort to automate trial and error of parallel processing area search, I propose automatic GPU offloading using evolutionary computation method.

Regarding applications placement, there is research on optimizing the inserted position of VN (Virtual Network) for a group of servers on the network as an effective use of network resources [33]. In [33], the optimum placement of VN is determined in consideration of communication traffic. However, it is for single-resource of virtual networks, with the aim of reducing carrier facility costs and overall response time, conditions such as the processing time of individual users' applications and requirements of the cost and response time are not taken into consideration.

### 2.2 Previous proposals

To adapt software to an environment, I previously proposed environment-adaptive software [28], the processing flow of which is shown in Figure 1. The environment-adaptive software is achieved with an environment-adaptation function, test-case database (DB), code-pattern DB, facility-resource DB, verification environment, and production environment.

Step 1: Code analysis
Step 2: Offloadable-part extraction
Step 3: Search for suitable offload parts
Step 4: Resource-amount adjustment
Step 5: Placement-location adjustment
Step 6: Execution-file placement and operation verification
Step 7: In-operation reconfiguration

In Steps 1-7, the processing flow conducts code conversion, resource-amount adjustment, placement-location adjustment, and in-operation reconfiguration for environment adaptation. However, only some of the steps can be selected. For example, if we only want to convert code for a GPU, FPGA, or many-core CPU, we only need to conduct Steps 1-3.

I will summarize this section. Because most offloading to heterogeneous devices is currently done manually, I proposed the concept of environment-adaptive software and automatic offloading to heterogeneous devices. However, after automatic conversion, the adjustment of the appropriate location of the offloaded application has not been examined (corresponding to Step 5). Therefore, in this paper, I will consider a method for efficiently arranging automatically converted applications by satisfying the user's cost and response time requirements.

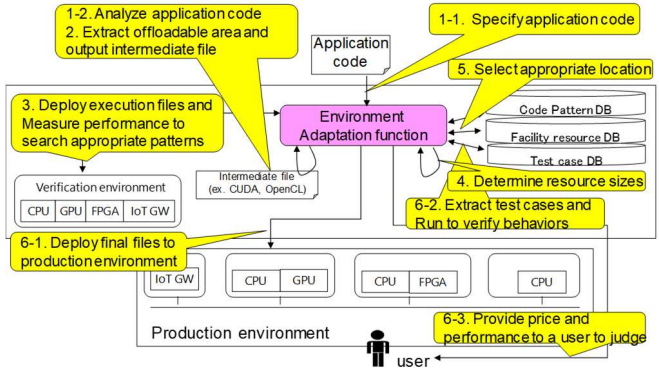

Fig. 1. Processing flow of environment adaptive software

## 3. Appropriate placement of applications

### 3.1 Consideration points to place applications

By methods such as [28], normal CPU programs are analysed using parsing library such as Clang [34] and can be automatically converted to offload devices such as GPUs using Genetic Algorithms (GA) [35]. In this subsection, I will consider placing the application in an appropriate location after the program conversion.

In method of [28] multiple offload patterns are repeatedly tried in the verification environment, and the appropriate offload pattern is selected. Therefore, the converted application is measured with the processing time, data amount, bandwidth used, calculation resource amount and so on in each offload pattern, including the finally selected offload pattern. Those measured values when each application is offloaded are used in the proper placement calculation.

Conventionally, applications have been placed in the cloud, data collected by IoT devices and the like has been transferred to a cloud server, where the data has been aggregated and analyzed. However, with the keywords edge computing and fog computing, there are increasing attempts to speed up the response time of applications by processing executions in the user environment or network edges that require real-time response.

In this paper as well, I will consider applications on the premise that they can be placed not only in the cloud but also at the network edge and user edge. However, at the network edge and user edge, the server concentration is lower than in the cloud and distributed, so the cost of computing resources is higher than in the cloud. Generally, the price of hardware such as CPU and GPU is constant regardless of the location. However, in a data center that operates the cloud, it is possible to monitor and control the air conditioning of the aggregated servers collectively, so the operating cost is cheaper.

For example, as a simple topology of the computational node link, Figure 2 can be considered. Fig. 2 shows the topology used in an example where an IoT device that collects data in a user environment sends data to the user edge, data is sent to the cloud via the network edge, and the analysis results are viewed by managers. Computational nodes are divided into three types: CPU, GPU, and FPGA. A node equipped with a GPU or FPGA also has a CPU, but it is provided as a GPU instance or an FPGA instance that includes CPU resources by virtualization technology such as NVIDIA vGPU [36] and MIG [37].

Applications are located on the cloud, network edge, and user edge, and the closer to the user environment, the lower the response time, but the higher the cost of computational resources. In this paper, I will deploy the converted application for GPU and FPGA, but when deploying, the user can make two types of requests. The first is a cost requirement, which specifies an acceptable price for operating the application, for example, operating it within 50 USD per month. The second is a response time request, which specifies an allowable response time when operating an application, for example, returning a response within 10 seconds.

In the conventional facility design, as considered in [33], for example, the location where the server accommodating the virtual network is placed is planned by analyzing at the long-term trend such as the amount of traffic increase.

On the other hand, this paper's target has two features. The first is that the application to be placed is not statically determined, but is automatically converted for GPU and FPGA, and the pattern suitable for the usage pattern is extracted through actual measurement through GA or other methods, so the application code and performance can change dynamically. For example, if the same Fourier transform program is used with large data for user A and small data for user B, the loop statements offloaded to the GPU might be different and it might be 10 times the performance compared to only CPU processing for A and 5 times the performance for B. Second, it is not only necessary to reduce carrier facility costs and overall response times, but it is also necessary to meet individual user requirements for costs and response times, and application placement policies can change dynamically.

Based on these two features, when there is a placement request from the user, the application placement in this paper is to sequentially place the converted applications on the server at that time to meet the user's request. If the cost performance does not improve even after converting the application, my method places applications without conversion. For example, a GPU instance costs twice as much as a CPU instance, and if converted application does not improve performance more than twice, it is better to place application without conversion. Also, if the computing resources and bandwidth have already been used up to the upper limit, they cannot be placed on that server.

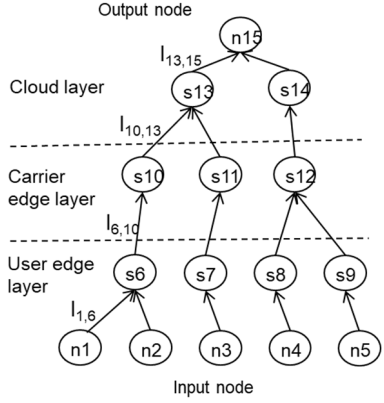

Fig. 2. Network topology example

### 3.2 Linear programming equations for appropriate application placement

In this subsection, I formulate a linear programming equation for calculating the appropriate location of the application. Figure 3 shows the equations and parameters.

Here, since the cost of devices and links, the upper limit of calculation resources, the upper limit of bandwidth depend on the server and network prepared by the operator, thus the parameter values are set in advance by the operator. The amount of computational resources, bandwidth, data capacity, and processing time used by the application when offloaded are determined by the measurements in the final selected offload pattern in the verification environment during automatic conversion. It is automatically set by the environment adaptation function.

The objective function and constraints change depending on whether the user request is a cost request or a response time request. If the request requires placement price within a month due to cost requirements, the minimization of the response time in (1) becomes the objective function, and the cost within (2) becomes one of the constraints. The constraint conditions that the resource upper limit of the server in (3) and (4) are not exceeded are also added. If the response time request requires the application to be placed within some seconds, the objective function is to minimize the cost of (5) corresponding to (2). The response time of (6) corresponding to (1) is one of the constraints within some seconds, and the constraints of (3) and (4) are also added.

(1) and (6) are equations for calculating the response time R_k of the application k, the objective function in the case of (1), and the constraint condition in the case of (6). (2) and (5) are equations for calculating the price P_k for operating the application k, the constraint condition in the case of (2), and the objective function in the case of (5). (3) and (4) are constraint conditions for setting the upper limit of the calculated resource and the communication band, which is calculated including the application placed by other users and checks the resource upper limit from being exceeded due to the application placement of the new user.

The linear programming equations (1)-(4) and (3)-(6) can be applied to conditions of network topology, conversion application type (cost and performance compared to only CPU processing), user requirements, and existing applications. By deriving a solution with a linear programming solver such as GLPK (Gnu Linear Programming Kit) or CPLEX (IBM Decision Optimization), an appropriate application location can be calculated. By sequentially performing the actual placement for plural users after the appropriate placement calculation, plural applications are placed based on the requests of each user.

$$R_k = \sum_{i \in Device}(A_{i,k}^d \cdot B_{i,k}^p) + \sum_{j \in Link}(A_{j,k}^l \cdot \frac{C_k}{B_k^l}) \quad (1)$$

$$\sum_{i \in Device} a_i(\frac{A_{i,k}^d \cdot B_k^d}{C_i^d}) + \sum_{j \in Link} b_j(\frac{A_{j,k}^l \cdot B_k^l}{C_j^l}) \leq P_k \quad (2)$$

$$\sum_{k \in App}(A_{i,k}^d \cdot B_k^d) \leq C_i^d \quad (3)$$

$$\sum_{k \in App}(A_{j,k}^l \cdot B_k^l) \leq C_j^l \quad (4)$$

$$P_k = \sum_{i \in Device} a_i(\frac{A_{i,k}^d \cdot B_k^d}{C_i^d}) + \sum_{j \in Link} b_j(\frac{A_{j,k}^l \cdot B_k^l}{C_j^l}) \quad (5)$$

$$\sum_{i \in Device}(A_{i,k}^d \cdot B_{i,k}^p) + \sum_{j \in Link}(A_{j,k}^l \cdot \frac{C_k}{B_k^l}) \leq R_k \quad (6)$$

$a_i$: Device usage cost
$b_j$: Link usage cost
$C_i^d$: Device calculation resource limit of #i
$C_j^l$: Link bandwidth limit of #j
$C_k$: Data size of #k application
$A_{i,k}^d$: Whether to use of #k application on #i device
$A_{j,k}^l$: Whether to use of #k application on #j link
$B_k^d$: Calculation resource of #k application
$B_k^l$: Bandwidth usage of #k application
$B_{i,k}^p$: Processing time of #k application on #i device

Fig. 3. The equations and parameters

# 4. Evaluation

## 4.1 Evaluation method

(a) Evaluated applications

The evaluated applications are the Fourier transform and image processing, which are expected to be used by many users.

The Fast Fourier Transform (FFT) is used in various situations of monitoring in IoT such as analysis of vibration frequency. NAS.FT [38] is one of the open source applications for FFT processing. It calculates the 2,048 * 2,048 size of the built-in sample test. When considering an application that transfers data from a device to a network in IoT, it is expected that the device will perform primary analysis such as FFT processing and send it in order to reduce network costs.

MRI-Q [39] computes a matrix Q, representing the scanner configuration for calibration, used in 3D MRI reconstruction algorithms in non-Cartesian space. In an IoT environment, image processing is often necessary for automatic monitoring from camera videos, and performance enhancements are requested in many cases. MRI-Q is a C language application and during application performance measurement, MRI-Q executes 3D MRI image processing to measure processing time using 64*64*64 size sample data.

From my previous automatic GPU and FPGA offloading methods [27][28], NAS.FT can be accelerated by GPU, and MRI-Q can be accelerated by FPGA. Performance improvement compared to CPU is 5 times and 7 times, respectively.

(b) Experiment conditions

The topology for arranging applications is composed of 3 layers as shown in Fig. 3, with 5 sites in cloud layers, 20 sites in carrier edge layers, 60 sites in user edge layers, and 300 input nodes. Assuming an application of IoT, IoT data is collected from the input node at the user edge, and analysis processing is performed at the user edge or carrier edge or cloud according to the response time requirements of the application.

All the servers to be analyzed are the assets held by a single operator, and the upper limit and price of the server and link are decided by the operator. In this evaluation experiment, the author decided on the following policy. As server numbers, there are 8 CPU servers, 4 GPU servers with 16GB RAM and 2 FPGA servers in the cloud, 4 CPU servers, 2 GPU servers with 8GB RAM and 1 FPGA server in the carrier edge, and 2 CPU servers and 1 GPU server with 4GB RAM in the user edge. Regarding server cost, for CPU, GPU, FPGA server, it is assumed that 6,000, 12,000, 14,400 USD will be collected in one year as the standard price of the servers in the cloud. When all resources of one server will be used (when using 16GB RAM for GPU server), the monthly fee is 500, 1,000, and 1,200 USD. Due to the aggregation effect, the monthly fee is 1.25 times and 1.5 times that of the cloud, assuming that the carrier edge and user edge will be expensive.

For links, a bandwidth of 100 Mbps is secured between the cloud and the carrier edge, and a bandwidth of 30 Mbps is secured between the carrier edge and the user edge. For the link cost, referring to the price of OCN Mobile One Full MVNO for IoT services (data transfer amount up to 500MB costs 5 USD per month, up to 1GB costs 8 USD per month, etc.), and I set prices that 100Mbps link fee is 80 USD per month, 30Mbps link fee is 50 USD per month.

As the resource used by the application, the value when actually offloaded to GPU or FPGA is used for the processing time. NAS.FT uses GPU 1GB RAM, usage band 2Mbps, transfer data amount 0.2MB, and processing time 5.8 seconds. MRI-Q uses 10 % of the FPGA server (the number of Flip Flop and Look UP Table used is the FPGA resource usage), the usage band is 1 Mbps, the transfer data amount is 0.15 MB, and the processing time is 2.0 seconds.

The experiment deploys up to 800 applications based on user requirements with set parameter values. The application is an IoT application and is supposed to analyze the data generated from the input node. Each input node generates placement requests randomly. Request number to place the applications is 800 times at a ratio of NAS.FT:MRI-Q = 3:1.

As a user request, a price condition or a response time condition is selected for each application when requesting placement. In the case of NAS.FT, the monthly upper limit of 70, 85 or 100 USD is selected for the price, and the 6, 7 or 10 second upper limit is selected for the response time. In the case of MRI-Q, the monthly upper limit of 125 or 200 USD is selected for the price, and the 4 or 8 second upper limit is selected for the response time. There are three patterns as variations of user requests.

Pattern 1 (P1): 6 types of requests by 1/6 for NAS.FT and 4 types of requests by 1/4 for MRI-Q.

Pattern 2 (P2): It selects the condition with the lowest price as the upper limit (initially 70 USD for NAS.FT, 125 USD for MRI-Q), and if there is no vacancy, the next cheapest price condition is selected.

Pattern 3 (P3): It selects the condition that the shortest time is the upper limit (initially 6 seconds for NAS.FT, 4 seconds for MRI-Q), and if there is no vacancy, the next shortest time condition is selected.

(c) Experimental tool

The placement is performed by a simulation experiment using the solver GLPK 5.0. It will be a simulation using tools to simulate a large-scale network layout. In actual use, when an application offload request comes in, an offload pattern is created by repeated performance tests using a verification environment, an appropriate amount of resources is determined based on the performance test results in the verification environment. Then appropriate placement is determined using GLPK or other solvers to meet user requests. After production placement, normal confirmation test and performance test are automatically performed, the result and price are presented to the user, and use is started after the user makes a judgment.

### 4.2 Results

Figure 4 is a graph in which the average price and the number of application placements are taken for three patterns, and Figure 5 is a graph in which the average response time and the number of application placements are taken for three patterns.

It was confirmed that pattern 2 fills in order from the cloud, and pattern 3 fills in order from the edge. In pattern 1, when various requests come in, they are arranged so as to satisfy the user requirements.

Regarding Fig. 4, in pattern 2, up to about 400 applications are placed in the cloud and the average price remains the lowest, but when the cloud is filled, it will gradually increase. In pattern 3, NAS.FT is placed from the user edge and MRI-Q is placed from the carrier edge, so the average price is high, and as it fills up, it is also placed in the cloud, so the average price becomes low. For pattern 1, the average price is in the middle of patterns 2 and 3 and is arranged according to the user's request, so the average price is appropriately reduced compared to pattern 3 which initially fills the edge.

Regarding Fig. 5, in pattern 2, up to about 400 applications are placed in the cloud and the average response time remains the highest, but when the cloud is filled, it gradually decreases. In pattern 3, NAS.FT is placed from the user edge and MRI-Q is placed from the carrier edge, so the average response time is the shortest, but as the number increases, it is also placed in the cloud, so the average response time becomes higher. Regarding pattern 1, the average response time is between patterns 2 and 3, and is arranged according to the user's request, so the average response time is appropriately reduced compared to pattern 2 which initially fills the cloud.

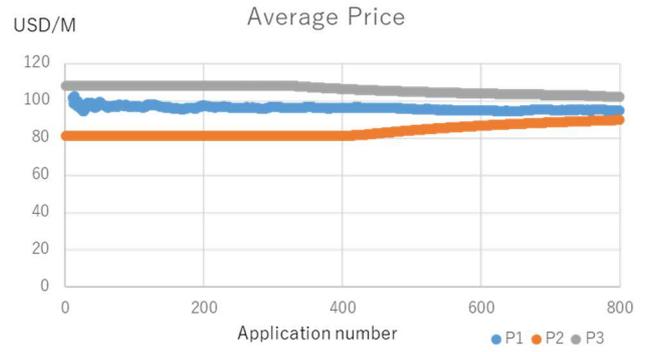

Fig. 4. Average price change with the number of application placements

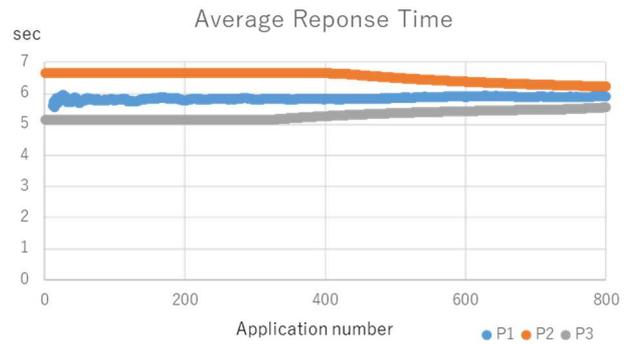

Fig. 5. Average response time change with the number of application placements

### 5. Conclusions

For a new element of my environment-adaptive software, in this paper, to respond to the user's cost request and response time request when automatically offloading to GPU or other devices, I proposed an application placement optimization method. Environment adaptive software adapts applications to the environments to use heterogeneous hardware such as GPUs and FPGAs appropriately.

The proposed method works after the program is converted and the amount of assigned resources is determined so that it can be processed by an offload device such as GPU. In the proposed method, first, the data capacity, the amount of calculation resources, the bandwidth, and the processing time of the application are set from the data of the performance test performed in the verification environment at the time of program conversion. Appropriate placement of applications is calculated based on the linear programming formula from the values set for each converted application and the values set by operators such as the cost of servers and links set in advance. When deploying an application, one is a constraint and the other is an objective function based on a user-specified price or response time request. An appropriate allocation is calculated by the linear programming solver, and the proposed method presents the price and so on to the user when the resource is allocated at the calculated location, and the production use is started after the user consents.

For the applications automatically offloaded to GPU and FPGA, the price condition and response time condition requested by the user, the number of application placements were changed, then the appropriate placement was calculated by the proposed method, and the effectiveness of the method was confirmed. In the future, I will consider not only calculating the proper placement at the beginning of actual use, but also reconfiguring when there is a more proper placement or converted offload pattern even during operation.

## References

(1) A. Putnam, et al., "A reconfigurable fabric for accelerating large-scale datacenter services," Proceedings of the 41th Annual International Symposium on Computer Architecture (ISCA'14), pp.13-24, June 2014.
(2) AWS EC2 web site, https://aws.amazon.com/ec2/instance-types/ Accessed 17 Dec. 2021.
(3) O. Sefraoui, et al., "OpenStack: toward an open-source solution for cloud computing," International Journal of Computer Applications, Vol.55, No.3, 2012.
(4) Y. Yamato, "Server Structure Proposal and Automatic Verification Technology on IaaS Cloud of Plural Type Servers," International Conference on Internet Studies (NETs2015), July 2015.
(5) Y. Yamato, et al., "Development of Service Control Server for Web-Telecom Coordination Service," IEEE International Conference on Web Services (ICWS 2008), pp.600-607, Sep. 2008.
(6) Y. Yamato, "Automatic system test technology of virtual machine software patch on IaaS cloud," IEEJ Transactions on Electrical and Electronic Engineering, Vol.10, Issue.S1, pp.165-167, Oct. 2015.
(7) Y. Yamato, "Proposal of Optimum Application Deployment Technology for Heterogeneous IaaS Cloud," 2016 6th International Workshop on Computer Science and Engineering (WCSE 2016), pp.34-37, June 2016.
(8) Y. Yamato, "Automatic verification for plural virtual machines patches," The 7th International Conference on Ubiquitous and Future Networks (ICUFN 2015), pp.837-838, July 2015.
(9) Y. Yamato, et al., "Fast Restoration Method of Virtual Resources on OpenStack," IEEE Consumer Communications and Networking Conference (CCNC2015), pp.607-608, Jan. 2015.
(10) Y. Yamato, et al., "Fast and Reliable Restoration Method of Virtual Resources on OpenStack," IEEE Transactions on Cloud Computing, DOI: 10.1109/TCC.2015.2481392, Sep. 2015.
(11) M. Hermann, et al., "Design Principles for Industrie 4.0 Scenarios," Rechnische Universitat Dortmund. 2015.
(12) Y. Yamato, "Proposal of Vital Data Analysis Platform using Wearable Sensor," 5th IIAE International Conference on Industrial Application Engineering 2017 (ICIAE2017), pp.138-143, Mar. 2017.
(13) Y. Yamato, et al., "Proposal of Real Time Predictive Maintenance Platform with 3D Printer for Business Vehicles," International Journal of Information and Electronics Engineering, Vol.6, No.5, pp.289-293, Sep. 2016.
(14) Y. Yamato and M. Takemoto, "Method of Service Template Generation on a Service Coordination Framework," 2nd International Symposium on Ubiquitous Computing Systems (UCS 2004), Nov. 2004.
(15) Y. Yamato, et al., "Security Camera Movie and ERP Data Matching System to Prevent Theft," IEEE Consumer Communications and Networking Conference (CCNC 2017), pp.1021-1022, Jan. 2017.
(16) Y. Yamato, et al., "Proposal of Shoplifting Prevention Service Using Image Analysis and ERP Check," IEEJ Transactions on Electrical and Electronic Engineering, Vol.12, Issue.S1, pp.141-145, June 2017.
(17) Y. Yamato, et al., "Analyzing Machine Noise for Real Time Maintenance," 2016 8th International Conference on Graphic and Image Processing (ICGIP 2016), Oct. 2016.
(18) Y. Yamato, "Experiments of posture estimation on vehicles using wearable acceleration sensors," The 3rd IEEE International Conference on Big Data Security on Cloud (BigDataSecurity 2017), pp.14-17, May 2017.
(19) P. C. Evans and M. Annunziata, "Industrial Internet: Pushing the Boundaries of Minds and Machines," Technical report of General Electric (GE), Nov. 2012.
(20) T. Sterling, et al., "High performance computing : modern systems and practices," Cambridge, MA : Morgan Kaufmann, ISBN 9780124202153, 2018.
(21) J. E. Stone, et al., "OpenCL: A parallel programming standard for heterogeneous computing systems," Computing in science & engineering, Vol.12, No.3, pp.66-73, 2010.
(22) J. Sanders and E. Kandrot, "CUDA by example : an introduction to general-purpose GPU programming," Addison-Wesley, 2011.
(23) J. Gosling, et al., "The Java language specification, third edition," Addison-Wesley, 2005. ISBN 0-321-24678-0.
(24) Y. Yamato, et al., "Automatic GPU Offloading Technology for Open IoT Environment," IEEE Internet of Things Journal, DOI: 10.1109/JIOT.2018.2872545, Sep. 2018.
(25) Y. Yamato, "Improvement Proposal of Automatic GPU Offloading Technology," The 8th International Conference on Information and Education Technology (ICIET 2020), pp.242-246, Mar. 2020.
(26) Y. Yamato, "Proposal of Automatic Offloading for Function Blocks of Applications," The 8th IIAE International Conference on Industrial Application Engineering 2020 (ICIAE 2020), pp.4-11, Mar. 2020.
(27) Y. Yamato, "Automatic Offloading Method of Loop Statements of Software to FPGA," International Journal of Parallel, Emergent and Distributed Systems, Taylor and Francis, DOI: 10.1080/17445760.2021.1916020, Apr. 2021.
(28) Y. Yamato, "Study of parallel processing area extraction and data transfer number reduction for automatic GPU offloading of IoT applications," Journal of Intelligent Information Systems, Springer, DOI:10.1007/s10844-019-00575-8, 2019.
(29) J. Fung and M. Steve, "Computer vision signal processing on graphics processing units," 2004 IEEE International Conference on Acoustics, Speech, and Signal Processing, Vol. 5, pp.93-96, 2004.
(30) S. Wienke, et al., "OpenACC-first experiences with real-world applications," Euro-Par 2012 Parallel Processing, pp.859-870, 2012.
(31) M. Wolfe, "Implementing the PGI accelerator model," ACM the 3rd Workshop on General-Purpose Computation on Graphics Processing Units, pp.43-50, Mar. 2010.
(32) E. Su, et al., "Compiler support of the workqueuing execution model for Intel SMP architectures," In Fourth European Workshop on OpenMP, Sep. 2002.
(33) C. C. Wang, et al., "Toward optimal resource allocation of virtualized network functions for hierarchical datacenters," IEEE Transactions on Network and Service Management, Vol.15, No.4, pp.1532-1544, 2018.
(34) Clang website, http://llvm.org/ Accessed 17 Dec. 2021.
(35) J. H. Holland, "Genetic algorithms," Scientific american, Vol.267, No.1, pp.66-73, 1992.
(36) NVIDIA vGPU software web site, https://docs.nvidia.com/grid/index.html Accessed 17 Dec. 2021.
(37) NVIDIA MIG website, https://docs.nvidia.com/datacenter/tesla/mig-user-guide Accessed 17 Dec. 2021.
(38) NAS.FT website, https://www.nas.nasa.gov/publications/npb.html Accessed 17 Dec. 2021.
(39) MRI-Q website, http://impact.crhc.illinois.edu/parboil/ Accessed 17 Dec. 2021.